\documentclass[twocolumn]{aastex63}

\usepackage{lineno}
\usepackage{amsmath}
\usepackage{amssymb}
\usepackage{graphicx}
\usepackage{color}
\usepackage{mathrsfs}
\usepackage{epsfig}
\usepackage{comment}
\usepackage{ulem}

\usepackage[graphicx]{realboxes}
\usepackage{booktabs,array}

\newcommand{\msun}{M_\odot}

\newcommand{\cc}{{\rm cm}^{-3}}

\newcommand{\msunyr}{M_\odot~{\rm yr}^{-1}}

\newcommand{\K}{{\rm K}}
\newcommand{\beq}{\begin{equation}}
\newcommand{\eeq}{\end{equation}}

\received{\today}

\shorttitle{Universal phase transition of AGNs}
\shortauthors{Inayoshi et al.}

\begin{document}

\title{Universal transition diagram from dormant to actively accreting supermassive black holes}

\correspondingauthor{Kohei Inayoshi, Kohei Ichikawa}
\email{inayoshi.pku@gmail.com, k.ichikawa@astr.tohoku.ac.jp}

\author[0000-0003-1451-6836]{Kohei Inayoshi}
\affil{Kavli Institute for Astronomy and Astrophysics, Peking University, Beijing 100871, China}

\author[0000-0002-4377-903X]{Kohei Ichikawa}
\affil{Astronomical Institute, Graduate School of Science
Tohoku University, 6-3 Aramaki, Aoba-ku, Sendai 980-8578, Japan}
\affil{Frontier Research Institute for Interdisciplinary Sciences, Tohoku University, Sendai 980-8578, Japan}

\author[0000-0003-1451-6836]{Luis C. Ho}
\affil{Kavli Institute for Astronomy and Astrophysics, Peking University, Beijing 100871, China}
\affil{Department of Astronomy, School of Physics, Peking University, Beijing 100871, China}

\begin{abstract}

The vast majority of supermassive black holes (SMBHs) in the local universe exhibit levels of activity much 
lower than those expected from gas supplying rates onto the galactic nuclei, and only a small fraction of 
silent SMBHs can turn into active galactic nuclei. 
Revisiting observational data of very nearby SMBHs whose gravitational spheres of influence are spatially reached by 
the \textit{Chandra} X-ray satellite, we find that the level of BH activity drastically increases from the quiescent phase 
when the inflow rate outside of the BH influence radius is higher than 0.1\% of the Eddington accretion rate. 
We also show that the relation between the nuclear luminosity and gas accretion rate from the BH influence 
radius measured from X-ray observations is well described by the universal state transition of accreting SMBHs, 
as predicted by recent hydrodynamical simulations with radiative cooling and BH feedback. 
After the state transition, young massive stars should form naturally in the nucleus, 
as observed in the case of the nearest SMBH, Sagittarius A$^\ast$, which is currently quiescent but was recently active.

\end{abstract}


\section{Introduction} \label{sec:intro}

Supermassive black holes (SMBHs) are ubiquitously harbored at the centers of massive nearby galaxies, 
and are believed to coevolve with their host galaxies through BH feeding and energetic feedback over 
cosmic time \citep[][hereafter KH13]{Kormendy_Ho_2013}.
Most SMBHs in the local universe are nearly quiescent, low-luminosity active galactic nuclei (AGNs) 
with bolometric luminosities $L_{\rm bol}/L_{\rm Edd}\ll 10^{-3}$, where $L_{\rm Edd}$ is the Eddington luminosity \citep{Ho_2008,Ho_2009}. 
The radiatively inefficient accretion flow (RIAF) model has been proposed to explain the nature of low-luminosity 
accreting SMBHs \citep{Ichimaru_1977,NY_1994,Stone_1999,Narayan_2000,Quataert_2000,Yuan_Narayan_2014}.
Two notable examples are Sagittarius A$^\ast$ (Sgr A$^\ast$) and the SMBH at the center of the giant 
elliptical galaxy NGC 4486 (M87). 
Recently, the Event Horizon Telescope project pulled together observatories around the globe and succeeded in 
imaging the accretion flow onto the M87 SMBH and presumably its black hole shadow \citep{EHT_1_2019,EHT_4_2019}.
By comparing the observed image to those obtained by ray-traced general relativistic magnetohydrodynamic simulations 
of BH accretion, our understanding of gas dynamics in the strong-field regime of general relativity will be 
improved significantly \citep[e.g.,][]{Dexter_2012,Moscibrodzka_2016,Ryan_2018}.

Observations of nearby SMBHs also provide a great opportunity to explore BH accretion dynamics at 
the BH gravitational influence radius (hereafter, Bondi radius), which is on the order of $\sim 1-100$ parsecs
($\gg$ the event horizon scales), from which mass inflows 
occur at the so-called Bondi accretion rate $\dot{M}_{\rm B}$.
This Bondi accretion rate is supposed to be an upper limit of the BH feeding rate and lead to radiative output with 
$L_{\rm bol}/L_{\rm Edd}\approx 10^{-3}$.
However, this level of energy release is far more luminous than actually observed, assuming a canonical value of 
10\% for the radiative efficiency \citep{SS_1973} or an even lower value ($\sim$ a few \%) for RIAFs 
\citep{Xie_Yuan_2012,Ryan_2017}.
The deficit of radiative luminosity suggests that only a fraction of inflowing matter from the Bondi scale 
actually feeds the BH and produces radiative/mechanical output that we observe in the nuclei of low-luminosity SMBHs.
Therefore, drawing the relation between the mass inflow rate and radiative output is essential to understand the 
mechanisms governing the fate of the fuel reservoir, and to obtain a consistent picture to link quantitatively 
dormant SMBHs and more luminous AGNs.

In this paper, we revisit observational data of very nearby SMBHs for which the spatial resolution of 
the \textit{Chandra} X-ray satellite reaches the Bondi scales, and report that the observational data clearly show the existence 
of a universal state transition for BH activity.
The level of radiative output before the state transition and critical mass inflow rate for the transition can be described well 
by recent numerical simulations taking into account radiative processes and BH feedback.

\section{Sample assembly}

High-spatial resolution X-ray observations using the \textit{Chandra} satellite give unique information on the properties 
of inflowing gas around the Bondi radius, where the gas is turned into hot and diffuse plasma with temperature 
$T \approx 0.2-1~{\rm keV}$ and electron density $n_{\rm e} \approx 0.1-10$~cm$^{-3}$.
Even with the excellent spatial resolution of $\sim 0.5^{\prime\prime}$, the Bondi scale, at most on the order of 
$\sim100$~pc, is resolvable only for a small number of sources at distances up to $50$~Mpc.

In this paper, we adopt the compilation by \cite{Pellegrini_2005}, which contains a total of 
50 X-ray-faint galaxies with X-ray luminosities $L_\mathrm{2-10}\lesssim 10^{42}$~erg~s$^{-1}$ in the 2--10 keV 
band, for which \textit{Chandra} data are publicly available as of 2005.
For 15 of these sources, the physical parameters of diffuse hot gas surrounding the nuclear regions, such as 
thermal temperature and electron density, have been directly measured with \textit{Chandra} observations.
We note that NGC 821, NGC 1553, and NGC 5128 are excluded from our sample because the electron densities 
for the first two sources in the literature are not observed values but just assumed to be $n_{\rm e}\simeq 0.1~\cc$,
and the density for the third source is not measured by \textit{Chandra} observations but \textit{XMM-Newton},
which has a lower spatial resolution of $\simeq 60^{\prime \prime}$.
In addition, we collect some interesting sources whose Bondi scales are reported to be resolved.
One is NGC 224 (M31), which is the nearest major galaxy to us and contains an SMBH in the center, and 
detailed \textit{Chandra} observations are available \citep{Dosaj_2002,Garcia_2005}.
The others are nearby galaxies that have been extensively studied: NGC 1332 \citep{Humphrey_2009}, 
NGC 3115 \citep{Wong_2014}, and NGC 1407 \citep{Humphrey_2006}. 
Moreover, \citet[hereafter R13]{Russell_2013} compiled \textit{Chandra} observations to discuss 
the Bondi accretion rates for the SMBHs in 13 nearby galaxies, providing us with eight new sources\footnote{ 
Note that NGC 4778, the brightest galaxy in HCG 62, is excluded because the BH mass measurement is 
not based on reliable methods (see \S\ref{sec:MBH}).}.
Finally, we have added four galaxies (NGC 315, NGC 2681, NGC 4278, and NGC 5005) with clear hot diffuse gas
 emission \citep{Younes_2011}.
In total, we have added 16 more sources, resulting in the total number of  31 sources.

In what follows, we describe the manner of observational data assembly for 
the 31 low-luminosity SMBHs.
The estimated physical quantities for those objects are summarized in Table \ref{tab:1}.

\subsection{Distance $(D)$}
It is crucial to obtain a homogeneous set of reliable distances for our sample.
Since most of our sample are very nearby sources, redshift-independent distances are used whenever possible.  
We follow the priorities for the choices of distance from KH13 (see their Tables 2 and 3).  
For 23 out of the 31 sources, we use distances based on surface brightness fluctuation measurements for individual 
galaxies in the Virgo and Fornax clusters \citep{Blakeslee_2009,Blakeslee_2010}.  
For NGC 1291, we adopt the distance based on the tip of the red giant branch \citep{McQuinn_2017}.
For Sgr A$^\ast$, we adopt the distance measured from resolved stellar dynamics \citep{Genzel_2010}.
Distances for the remaining six sources are taken from the mean values of several distance determinations listed in NED, 
mainly based on Cepheid variables, surface brightness fluctuations, tip of the red giant branch, and RR Lyrae stars.

\subsection{Black Hole Mass $(M_\bullet)$}
\label{sec:MBH}
We collect the BH masses of 16 sources based on spatially resolved stellar or ionized gas kinematic observations, 
as compiled in KH13 and the references therein.
Since the other 15 sources do not have reliable dynamical mass measurements, we estimate their BH masses 
using the empirical relation given in Eq. (7) of KH13,
\begin{equation}
M_\bullet = 3.09^{+0.37}_{-0.33} \times 10^8 \left(\frac{\sigma}{200~ \mathrm{km~s^{-1}}}\right)^{4.38 \pm 0.29} M_\odot,
\end{equation}
where $\sigma$ is the central velocity dispersion of the bulge stars, taken from the Hyperleda database \citep{Paturel_2003}. 
The mean error of the BH masses for the sample is $\Delta M_\bullet/M_\bullet=0.26$.

\subsection{Bolometric Luminosity $(L_\mathrm{bol})$}

Ideally, the AGN bolometric luminosities should be measured directly from their broadband spectral energy distributions (SEDs).  
In the case of Sgr A$^\ast$, an almost complete nuclear SED is available from the radio to X-ray band \citep{Narayan_1998}, 
integration of which yields $L_\mathrm{bol}=1.2 \times 10^{36}$~erg~s$^{-1}$.

Unfortunately, complete broadband nuclear SEDs are not available for most of our sources.  
One of the most secured indirect methods is to estimate $L_\mathrm{bol}$ from the 2--10~keV absorption-corrected X-ray 
luminosity ($L_{2-10}$) using a bolometric correction of $C_{\rm X}\equiv L_{\rm bol}/ L_{\rm 2-10} = 15.8$ \citep{Ho_2009}.  
Since our targets are very faint, the X-ray data must be obtained with spatial resolution high enough to distinguish the nuclear 
source from the host galaxy.  
We collect the X-ray luminosities from \textit{Chandra} observations achieving the best spatial resolution in the X-ray band.
Note that there are six sources that have no X-ray core coincident with the assumed optical nucleus, have no power-law component
in their nuclear spectra, or have a significant level of contamination from X-ray binaries (see Table~\ref{tab:1}).  
We treat their nuclear X-ray luminosities as upper limits in Figure \ref{fig:summary}.

It is quite difficult to evaluate the uncertainties of the estimated bolometric luminosities.  
A major concern is the possible underestimation of intrinsic absorption of the emergent X-ray radiation.  
Low-luminosity AGNs, however, generally suffer from very minimal intrinsic absorption \citep{Ho_2008}, and this effect typically 
introduces an uncertainty only at the level of a factor of 2 to our luminosity estimates \citep{Ho_2009}.  
Accordingly, we assign errors of 50\% to $L_{\rm bol}$.

The Eddington ratio is defined by $\lambda_\mathrm{Edd} \equiv  L_\mathrm{bol} / L_\mathrm{Edd}$, 
where $L_\mathrm{Edd}= 1.26 \times 10^{38} (M_\bullet / M_\odot)$~erg~s$^{-1}$. 
We estimate the uncertainty of $\lambda_\mathrm{Edd}$ by propagating the errors on $L_\mathrm{bol}$ 
and $M_\bullet$.

\subsection{Temperature $(T)$ and Electron Density $(n_\mathrm{e})$}
Both the temperature and electron density are key parameters needed to estimate the Bondi radius and 
the Bondi accretion rate.  
For each object in our sample, we select the deepest \textit{Chandra} observations from the literature 
to obtain $T$ and $n_\mathrm{e}$ of the diffuse hot gas in the nuclear region.
Although the details of the analysis differ from study to study, the general methodology to estimate these 
two quantities follows a similar manner.  
Whenever possible, the diffuse component is extracted after removing point sources arising from X-ray binaries 
and the central nuclear component.
Since the nuclear component usually has a characteristic power-law spectrum in the hard X-ray band above 
$\sim 1$~keV, this fact enables us to distinguish the nuclear emission from the diffuse emission that has a peak 
at $\lesssim 1$~keV.
Therefore, contamination from the nuclear component is not a serious issue for obtaining the temperature of 
diffuse hot gas surrounding the SMBH.

The spatially resolved region containing diffuse gas is divided into radial annuli with 
sufficient counts for spectral fitting \citep[see a typical example in][]{Wong_2014}.  
The MEKAL \citep{Mewe_1985} and APEC \citep{Smith_2001} models are commonly used for spectral fitting, 
and both are implemented in the standard X-ray fitting tool XSPEC \citep{Arnaud_1996}.  
The model fit yields the temperature $T$ from the peak of the thermal bump at $E=0.3-1$~keV, 
and the electron density $n_\mathrm{e}$ from the spectral normalization, producing radial profiles of 
the two quantities.
The literature does not always provide the uncertainties for estimating $T$ and $n_\mathrm{e}$.  
In such cases, we adopt the mean error of the sources with robust error estimates; 
$\Delta T/T = 0.10$ and $\Delta n_\mathrm{e}/n_\mathrm{e}=0.43$.

In Table~\ref{tab:1}, we list the temperatures and electron densities measured either at $r=R_{\rm B}$ or 
at the inner-most radii that the {\it Chandra} observations can reach if the Bondi radii are not resolved. 
As exceptions, however, we present the electron densities extrapolated to the Bondi radii 
of seven unresolved sources studied in R13 (NGC 507, NGC 1316, NGC 4696, NGC 5044, NGC 5813, 
NGC 5846, and NGC 6166), for which the density profiles are fitted with three different models 
(a power-law model continuing a steep density gradient to $R_{\rm B}$, a $\beta$-model flattening to 
a constant, and a shallow S\'ersic profile with $n = 4$) with errors arising from the different model 
assumptions.

\subsection{Bondi Radius $(R_{\rm B})$ and Accretion Rate $(\dot{M}_{\rm B})$}

The Bondi radius is the characteristic radius within which the gravitational force by the central BH dominates 
the thermal pressure gradient force of the gas surrounding the BH \citep{Bondi_Hoyle_1944, Bondi_1952}, defined by
\begin{equation}\label{Eq:Rbondi}
R_{\rm B}\equiv \frac{GM_\bullet}{c_{\rm s}^2}
\simeq 1.68~M_8 T_\mathrm{keV}^{-1}~\mathrm{pc},
\end{equation}
where $M_8=M_\bullet/(10^8~\msun)$ and $T_\mathrm{keV}=kT/\mathrm{keV}$.
We emphasize that our definition of $R_\mathrm{B}$ is smaller by a factor of 2 compared to the one in 
\cite{Pellegrini_2005}, which uses the diameter instead of the radius.
Note that the uncertainties in estimating the Bondi radius are mainly due to the uncertainties from the BH mass 
measurements, and the mean uncertainty is $\Delta R_\mathrm{B}/R_\mathrm{B} \approx 0.3$.

We divide the sample into two groups based on whether or not the angular size of the Bondi radius 
($\theta_{\rm B}\equiv R_{\rm B}/D$) within the estimated uncertainty is larger than $0.5^{\prime\prime}$, 
which is the best achievable resolution by \textit{Chandra}.
Based on this criterion, we define sources having sufficiently large Bondi radii 
($\theta_\mathrm{B}\geq 0.5^{\prime\prime}$) as resolved (see the first 11 sources labeled as ``Bondi = Y'' in Table~\ref{tab:1}).
Note that the Bondi radius of the BH in NGC 4594 might be marginally resolved within the error of the BH mass 
measurement, but the spatial resolution of its \textit{Chandra} observation reaches only 
$\simeq 2.5^{\prime\prime}$ \citep{Pellegrini_2003}.  
Thus, we consider this source to be unresolved (see the remaining 20 sources labeled as ``Bondi = N'' in Table~\ref{tab:1}).

The Bondi accretion rate is defined as the mass inflow rate through the Bondi radius,
\begin{eqnarray}
\dot{M}_{\rm B} &\equiv& 4\pi q(\gamma) \rho
\frac{G^2M_\bullet ^2}{c_{\rm s}^3}\nonumber\\
&\simeq &7.0 \times 10^{-6}~n_{\rm e,0.1} M_8^2 T_\mathrm{keV}^{-3/2}~\msunyr,
\end{eqnarray}
where $q(\gamma)=1/4$ for $\gamma =5/3$ and $n_{\rm e,0.1}=n_{\rm e}/(0.1~\mathrm{cm}^{-3})$.
The Bondi accretion rate normalized by the Eddington accretion rate is given as
\begin{equation}
\dot{m}_{\rm B}\equiv \frac{\dot{M}_{\rm B}}{\dot{M}_{\rm Edd}}
\simeq 3.1 \times 10^{-6}~ n_{\rm e,0.1} M_8 T_\mathrm{keV}^{-3/2}.
\end{equation}
For the unresolved sources ($\theta_{\rm B} < 0.5^{\prime\prime}$), their Bondi accretion rates 
are estimated by extrapolating the electron densities at $R_{\rm B}$ from the measured values, namely by multiplying 
by a factor of $f~(\equiv 0.5^{\prime\prime}/\theta _{\rm B}>1)$ with an assumed density profile of 
$n_{\rm e} \propto r^{-1}$.
The choice of the slope $\beta$ is consistent with the profiles for two well-resolved objects: 
$\beta = -1.0\pm0.2$ for NGC 4486 (M87) \citep{Russell_2015} and $\beta = -1.05\pm0.25$ for 
NGC~3115 \citep{Wong_2014}. 
Note that the extrapolation has been taken into account for the seven unresolved sources studied by R13.

For NGC 1291 and NGC 4594, the extrapolated values of electron density are treated as lower limits 
because their nuclear regions are poorly resolved ($\theta \simeq 2.9^{\prime\prime}$ and $2.5^{\prime\prime}$,
respectively; \citealt{Irwin_2002,Pellegrini_2003}).
For NGC 221 (M 32), the electron density is the volume-averaged value within 30$^{\prime\prime}$ and is thus treated 
as a lower limit \citep{Ho_2003}.
For convenience, we list the ``uncorrected'' Bondi accretion rate and the correction factor separately in Table~\ref{tab:1}, 
but show the ``corrected'' Bondi accretion rate in Figure \ref{fig:summary}.
It is also worth noting that removing these three galaxies and five objects with no detection of nuclear X-ray emission from 
the unresolved sample, all of the other 12 unresolved sources are located in the transition region in Figure \ref{fig:summary} within the errors.

\cite{Younes_2011} did not provide $n_\mathrm{e}$ for the four low-luminosity sources included in their study, 
only estimates of the emission measure for which the volume information is integrated.  
To derive the density from the emission measure, we assume that the volume contribution to the emission measure 
is inside a radius of $1^{\prime\prime}$ and that the slope of the density profile is $\beta=-1$.  
The choice of radius is quite arbitrary, but the estimated value for NGC~315 ($n_\mathrm{e}=0.36$~cm$^{-3}$) 
agrees well with the density reported by other studies \citep{Worrall_2003}.

\section{Results}

\begin{figure*}
\begin{center}
\includegraphics[width=125mm]{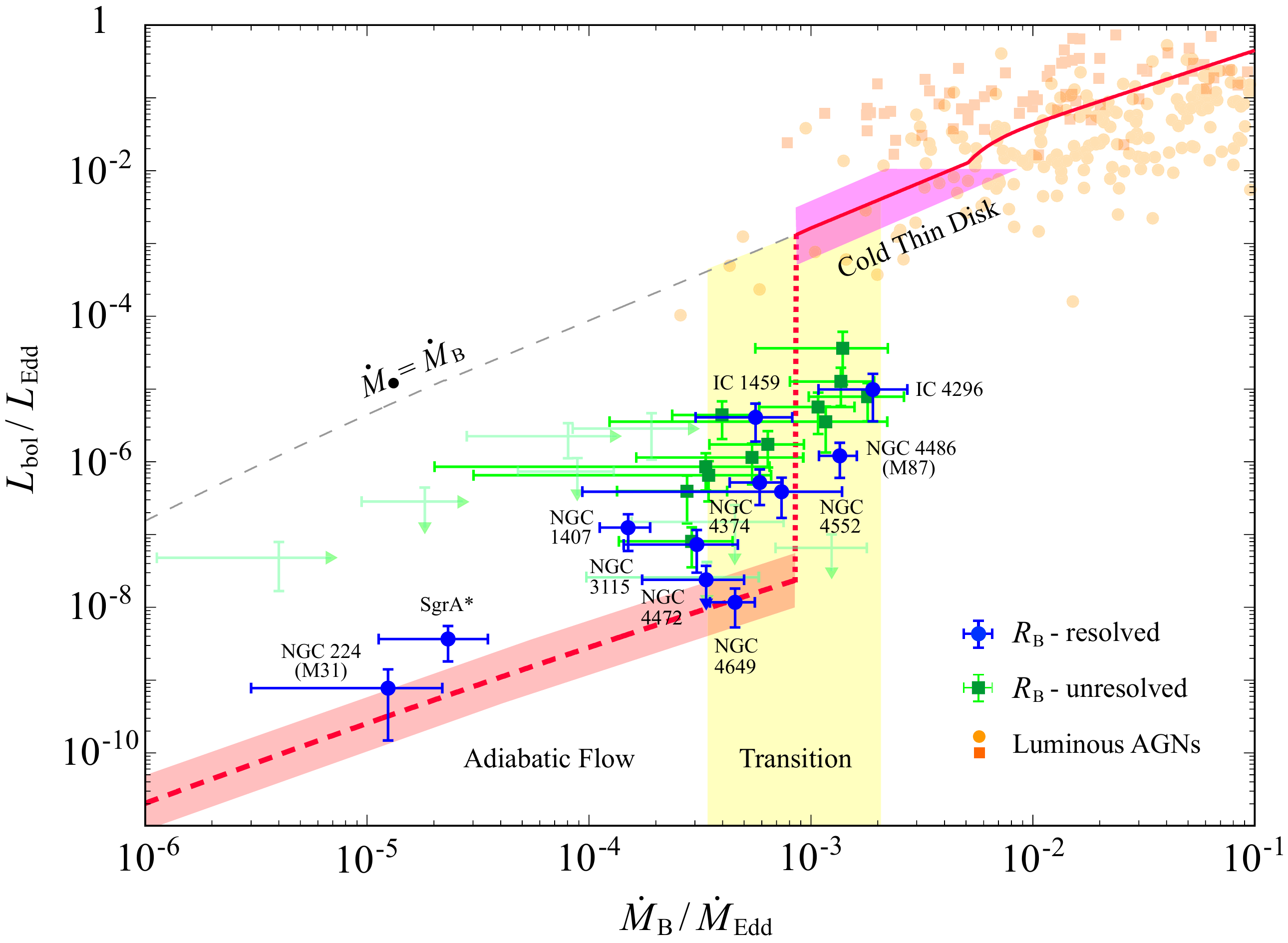}
\caption{Radiative luminosity and Bondi accretion rate for nearby quiescent and low-luminosity accreting SMBHs 
whose Bondi scales are resolved (blue) and unresolved (green) by {\it Chandra}\ observations.
Numerical simulation results are shown by the red solid and dashed curves.
For lower values of $\dot{M}_{\rm B}/\dot{M}_{\rm Edd}$ ($<10^{-3}$), the radiative luminosity is significantly reduced 
from that for $\dot{M}_\bullet =\dot{M}_{\rm B}$ due to the suppression of the BH feeding by turbulent gas motion 
in the disk (red region). 
The transition from RIAFs to cold accretion disks occurs at 
$4\times 10^{-4}<\dot{M}_{\rm B}/\dot{M}_{\rm Edd}< 2\times 10^{-3}$ (yellow shaded region). 
After the transition, the cold thin disk tends to be unstable to its self-gravity (magenta region).
These observational and theoretical results bridge the gap between the dormant SMBH population and the faint 
end of the Seyfert galaxy population in the local universe.
Orange circles and squares represent AGNs from the {\it Swift}/BAT AGN and PG quasar catalog, respectively. }
\label{fig:summary}
\end{center}
\end{figure*}

Figure \ref{fig:summary} presents the relation between the luminosity and Bondi accretion rate for the samples whose Bondi scales 
are resolved (blue) and unresolved (green).
The luminosity and accretion rate are both normalized by their Eddington values, as BH accretion systems are 
characterized by these two dimensionless quantities, not by the actual physical scales.
Assuming that the BH feeding rate is equal to the Bondi accretion rate, as in the original advection-dominated accretion 
flow model (gray thin dashed line), the radiative luminosity is expected to be much higher than the observed values by 
several orders of magnitude. 
While the luminosity discrepancy is large at lower Bondi rates ($\dot{M}_{\rm B}\ll 10^{-3}~\dot{M}_{\rm Edd}$), 
the luminosity increases dramatically at $\dot{M}_{\rm B}\simeq 10^{-3}~\dot{M}_{\rm Edd}$ and catches up to 
the lowest levels of the activity observed in low-luminosity AGNs ($L_{\rm bol}\approx 10^{-4}~L_{\rm Edd}$; orange symbols). 
Even in the face of uncertainties of electron density measurements for several unresolved objects, 
the overall trend covering many orders of magnitude in the phase diagram of Figure \ref{fig:summary} is seen robustly.

\begin{figure}
\begin{center}
\includegraphics[width=80mm]{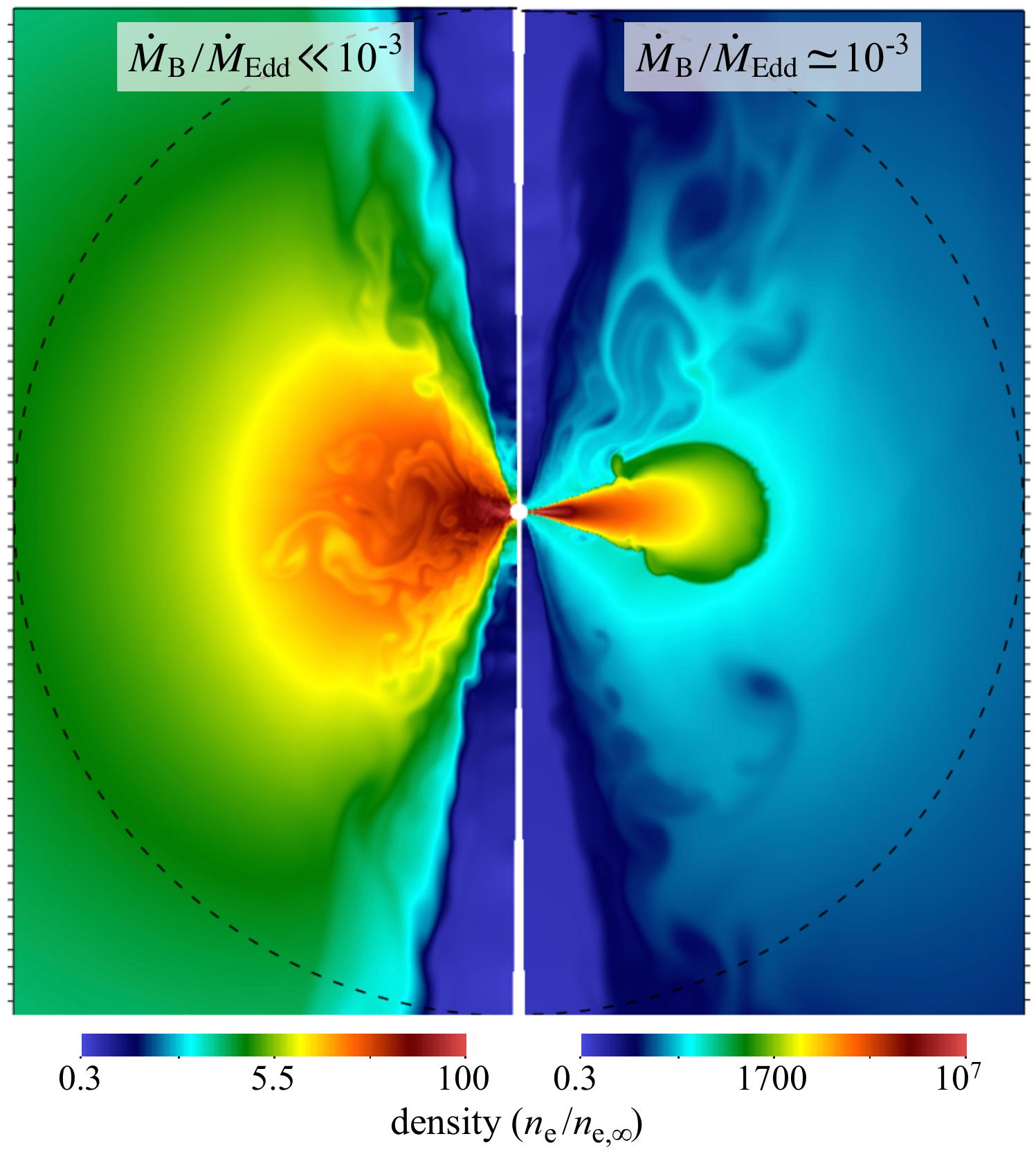}
\caption{Density distribution of accretion flows onto an SMBH accreting at two different rates,
from the edge-on view (i.e., the horizontal and vertical axes are parallel and perpendicular
to the equator of the accretion flow).  
For lower accretion rates ($\dot{M}_{\rm B} \ll 10^{-3}~\dot{M}_{\rm Edd}$; left), 
the RIAF is kept adiabatic and highly turbulent, forming 
a geometrically thick disk inside the Bondi radius (dashed line).
For the higher accretion rates ($\dot{M}_{\rm B} \simeq 10^{-3}~\dot{M}_{\rm Edd}$; right), the accreting gas begins to cool, 
collapses toward the equator, and forms a cold, geometrically thin disk.}
\label{fig:cont}
\end{center}
\end{figure}

Figure \ref{fig:cont} shows the two-dimensional distribution of the density distribution of rotating accretion flows onto an SMBH, 
obtained from two-dimensional hydrodynamical simulations with radiative processes and BH feedback \citep{Inayoshi_2019},
where feedback is modeled with the prescription obtained by \citet{Sazonov_2005} without injecting disk winds (see also \citealt{Yuan_2018})\footnote{
Several previous works proposed the importance of mass loss via winds to explain the nature of low-luminosity AGNs
\citep[e.g.,][]{Blandford_Begelman_1999,LOS_2013,Bu_2016}}.
Unlike previous studies that have investigated accretion flows on smaller scales assuming a compact and gravitationally 
bound disk as the initial state, these numerical simulations focus on the accretion dynamics at larger scales covering 
the Bondi radius (dashed circles), where the gas is weakly bound initially.
This type of simulation allows us to set plausible initial and boundary conditions, which can be directly measured 
in the nuclear regions surrounding quiescent SMBHs. 
For the lower density case with $\dot{M}_{\rm B}/\dot{M}_{\rm Edd}\ll 10^{-3}$ (left panel), radiative cooling is inefficient and 
viscous energy dissipation heats the gas, leading to a hot and turbulent accretion flow through a geometrically thick disk.
Importantly, the inflow rate decreases toward the center as $\dot{M}_{\rm inf}(r)\simeq \dot{M}_{\rm B}\cdot (r/R_{\rm B})$ 
due to turbulent motion and results in a low value onto the central BH \citep{Quataert_2000,Inayoshi_2018}.
In such a hot accretion flow, thermal conductivity of electrons, instead of turbulence, transports energy outward at 
the inner region of $r< 10^{-2}~R_{\rm B}$.
Since suppression of the accretion by turbulence ceases in the inner region, the final BH feeding rate is 
\begin{equation}
\frac{\dot{M}_\bullet}{\dot{M}_{\rm Edd}} \simeq 1.5\times 10^{-6}~T_7^{-4/5}
\alpha_{-2}^{0.37} 
f_{-1}^{2/5}
\left(\frac{\dot{M}_{\rm B}}{10^{-3}~\dot{M}_{\rm Edd}}\right)^{3/5},
\end{equation}
where $T_7\equiv T/10^7~{\rm K}$ is the gas temperature, 
$\alpha_{-2}\equiv \alpha/0.01$ is the strength of viscosity, and $f_{-1}\equiv f_{\rm c}/0.1$ is the conductivity suppression factor 
due to magnetic fields \citep{Narayan_Medvedev_2001}.
In other words, the BH feeding rate is reduced by $2-3$ orders of magnitude from the Bondi accretion rate. 
As a result, the radiative luminosity reduces to
\begin{equation}
\frac{L_{\rm bol}}{L_{\rm Edd}} \simeq 3\times 10^{-8}
~T_7^{-4/3} \alpha_{-2}^{0.63} 
f_{-1}^{0.68}
\left(\frac{\dot{M}_{\rm B}}{10^{-3}~\dot{M}_{\rm Edd}}\right),
\end{equation}
where the radiative efficiency model for a hot accretion disk is adopted (\citealt{Ryan_2017}; see also Figure 2 in \citealt{Inayoshi_2019}).
The level of radiative luminosity agrees well with the observational results for SMBHs accreting at 
$\dot{M}_{\rm B}/\dot{M}_{\rm Edd}\ll 10^{-3}$, as shown in Figure \ref{fig:summary} (red region).
Note that the width of the red region reflects the uncertainties of the conductivity suppression factor, which has a range of
$0.03\leq f_{\rm c} \leq 0.3$.

By contrast, for the highest rate of $\dot{M}_{\rm B}/\dot{M}_{\rm Edd}\gtrsim 10^{-3}$ the accreting gas collapses to 
the midplane and forms a geometrically thin disk because of radiative cooling (right panel).  
Since the entropy generated by 
viscosity is radiated away, turbulent gas motion ceases, and the gas accretion rate through the disk becomes as high 
as $\dot{M}_\bullet \simeq \dot{M}_{\rm B}$. 
This cooling transition dramatically increases the luminosity by several orders of magnitude (Figure \ref{fig:summary}, yellow region).
We note that this critical rate required for the cooling transition around the BH influence radius, 
$\dot{M}_{\rm B}\gtrsim 10^{-3}~\dot{M}_{\rm Edd}$ \citep{Gaspari_2015,Inayoshi_2019} is $1-2$ orders of magnitude lower than 
the critical rate for a compact accretion disk near the BH \citep{Yuan_2001,Yuan_2003},
the latter of which would be more relevant to the disk state transition for X-ray binaries.

\section{Discussion}

\subsection{Relationship with local bright AGNs}

Among the sample, 15 objects (five resolved and 10 unresolved ones) are located in the transition region of 
Figure \ref{fig:summary} and are as luminous as $10^{-7}<L_{\rm bol}/L_{\rm Edd}<10^{-4}$.
Those SMBHs bridge the gap between the adiabatic accretion flows and the faint end of Seyfert galaxies 
in the local universe \citep{Ichikawa_2019}.

In order to extend the state transition diagram to the brighter end, we consider two well-defined and complementary samples 
of local AGNs selected by X-rays and UV/optical surveys (orange circles and squares, respectively).
The X-ray-selected sample was derived from the all-sky 70-month \textit{Swift}/Burst Alert Telescope (BAT; \citealt{Baumgartner_2013}), 
whose $3-500~\mu$m infrared (IR) properties have been investigated \citep{Melendez_2014,Ichikawa_2017}
and quantified systematically from decomposition of the IR SED into the AGN and host galaxy components, 
i.e., star-formation activity \citep{Ichikawa_2019}.
The UV/optical-selected sample is based on the Palomar-Green (PG) quasar survey \citep{Boroson_1992}, and its 
star formation activity has also been investigated with decomposition of its IR SEDs \citep{Shangguan_2018}.
The rich multi-wavelength data sets of these objects enable us to estimate the bolometric luminosity and BH mass 
(i.e., Eddington luminosity and accretion rate), and also to study the statistical properties of their star formation 
rates (SFRs) on the host galaxy scales with the empirical relation \citep{Kennicutt_1998_ARAA},
\begin{equation}
{\rm SFR}_{\rm global}\simeq 0.45~\msunyr \left( \frac{L_{\rm FIR}}{{10^{43}~\rm erg~s}^{-1}}\right).
\end{equation}
We restrict the sample to bright objects with far-infrared (FIR) luminosities $L_{\rm FIR} > 10^{10}~L_\odot$ 
because the empirical relation is valid for star-forming galaxies and the IR SED fitting model is based on SED 
templates of galaxies with FIR luminosities above this threshold.

We note that because of their large distances, the Bondi radii for the \textit{Swift}/BAT and PG quasar samples 
are not resolved even with \textit{Chandra}, and thus their Bondi accretion rates are not estimated properly.
For the purpose of illustration in Figure \ref{fig:summary}, we replace the Bondi accretion rates on the horizontal axis 
with $0.01~{\rm SFR}_{\rm global}$ for all the 426 luminous AGNs.
This factor of 0.01 is arbitrary but may be of a similar order of magnitude as the Bondi accretion rate for the following reasons.
First, local Seyfert galaxies with high-angular resolution ($\sim 0.4-0.8^{\prime\prime}$) mid-infrared spectroscopy 
indicate that the SFRs in the nuclear regions ($<70$ pc) are, on average, 5 times lower than those measured in the 
circumnuclear regions on kiloparsec scales \citep{Esquej_2014}.
Second, numerical simulations studying the structure of circumnuclear disks suggest that strong star formation 
activity leads to turbulence in the disk, and turbulent viscosity efficiently induces mass accretion onto the Bondi 
scale at a rate of 10\% of the nuclear SFR \citep{Wada_Norman_2002,Inayoshi_2019}.
With these simple assumptions, the bright AGN population seems to follow the red curve in Figure \ref{fig:summary}, 
which assumes $\dot{M}_\bullet \simeq \dot{M}_{\rm B}$ with the radiative efficiency of a geometrically thin accretion disk.

\subsection{Past Activity of Sgr A$^\ast$}

Sgr A$^\ast$ is the SMBH in the center of the Milky Way with a mass of $M_\bullet \simeq 4.4\times 10^6~M_\odot$, 
whose activity is known to be very quiescent at present ($L_{\rm bol}/L_{\rm Edd}\lesssim 10^{-8}$).
However, several lines of observational evidence suggest that its past AGN activity was higher
 \citep{Kaifu_1972,Koyama_1996,Bland-Hawthorn_2003,Totani_2006,Ryu_2013}.
One of the most striking clues is the discovery of the Fermi bubbles, which are expanding above and below the Galactic 
plane with an age of roughly a few Myr \citep{Su_2010}. 
A short episode of AGN activity lasting $\sim O({\rm Myr})$ and injecting a total energy of order $\sim 10^{55}$ erg are 
required to create the bubbles.
This level of energy injection is achieved by assuming that the past AGN luminosity was 
$L_{\rm X}\simeq 10^{40}~{\rm erg~s}^{-1}$ ($L_{\rm bol}/L_{\rm Edd}\simeq 3\times 10^{-4}$, adopting a typical value 
of the bolometric correction for low-luminosity AGNs).
The Bondi accretion rate required to explain the energy output is on the order of 
$\dot{M}_{\rm B} \approx 10^{-3}~\dot{M}_{\rm Edd}~(\approx 10^{-4}~\msunyr)$, which is presumably
the accretion rate through the cold disk (i.e., $\dot{M}_{\rm d}\sim \dot{M}_{\rm B}$). 
In fact, the accretion rate is high enough for the disk to fragment into clumps and form stars by a spiral-mode 
gravitational instability, which is characterized by the Toomre parameter \citep{Toomre_1964}:
\begin{eqnarray}
Q &\simeq & \frac{3\alpha_{\rm eff} c_{\rm s}^3}{G\dot{M}_{\rm d}}\nonumber\\
&\simeq & 0.83
\left(\frac{\alpha_{\rm eff}}{0.03}\right)
\left(\frac{T_{\rm c}}{300~\K}\right)^{3/2}
\left(\frac{\dot{M}_{\rm d}}{\dot{M}_{\rm B}}\right)^{-1}
\lesssim 1,
\end{eqnarray}
where $\alpha_{\rm eff}~(\gtrsim 0.01)$ is the effective viscous parameter caused by the spiral arms and 
$c_{\rm s}~(\propto T_{\rm c}^{1/2})$ is the sound speed of cold gas with a temperature of $T_{\rm c}$.
The parameter range where the accretion disk would be gravitationally unstable after the transition is indicated 
with the magenta region in Figure \ref{fig:summary} (\citealt{Inayoshi_2019}, see also \citealt{Menou_Quataert_2001}).
This process naturally explains the existence of young massive stars in a thin stellar disk in the Galactic center \citep{Levin_Beloborodov_2003}.
Major episodes of BH accretion with $\gtrsim 0.01~L_{\rm Edd}$ and star formation would likely blow away the accreting 
gas and quench the activity of Sgr A$^\ast$.

\acknowledgments

We greatly thank Jeremiah P. Ostriker for useful discussions.
This work is partially supported by the National Science Foundation of China (11721303, 11991052, 11950410493), the National Key R\&D Program of China (2016YFA0400702), the Program for Establishing a Consortium for the Development of Human Resources in Science and Technology, Japan Science and Technology Agency, and the Japan Society for the Promotion of Science (JSPS) KAKENHI (18K13584). Numerical computations were carried out with High-performance Computing Platform of Peking University and Cray XC50 at the Center for Computational Astrophysics of the National Astronomical Observatory of Japan.


\bibliographystyle{aasjournal}

\begin{longrotatetable}
\begin{deluxetable*}{lccccccccccc}
\tablenum{1}
\tablecaption{Physical quantities of low-luminosity SMBHs in nearby galaxies.}
\tabletypesize{\scriptsize}
\tablewidth{0pt}
\tablehead{
\colhead{Name} & \colhead{$D$ (Mpc)} & \colhead{$\log (M_\bullet/\msun)$} & \colhead{$T$ (K)} &
\colhead{$n_\mathrm{e}$ ($\cc$)} & \colhead{$R_\mathrm{B}$ (pc)} & \colhead{$\log \dot{m}_\mathrm{B}$} &
\colhead{$\log \lambda_\mathrm{Edd}$} & \colhead{$\theta_\mathrm{B}$ ($^{\prime\prime}$)} & 
\colhead{$f$} & \colhead{Bondi} & \colhead{Reference} 
}
\startdata
NGC 224 & $0.784$ & $8.16 \pm 0.28$ & $0.50 \pm 0.04$ & $0.10 \pm 0.04$ & $4.8 \pm 3.1$ & $-4.91 \pm 0.33$ & $-9.11 \pm 0.35$ & $1.3 \pm 0.8$ & -- & Y & 
1, 2, 3, 4
\\
NGC 1407 & $29.0$ & $9.67 \pm 0.07$ & $0.780 \pm 0.020$ & $0.072 \pm 0.014$ & $100 \pm 16$ & $-3.82 \pm 0.11$ & $-6.90 \pm 0.23$ & $0.72 \pm 0.11$ & --  & Y & 
5, 6, 7, 8
\\
NGC 3115 & $9.54$ & $8.95 \pm 0.13$ & $0.37 \pm 0.11$ & $0.25 \pm 0.10$ & $41 \pm 17$ & $-3.51 \pm 0.23$ & $-7.14 \pm 0.26$ & $0.9 \pm 0.4$ & -- & Y & 
5, 9, 10, 11
\\
NGC 4374 & $18.51$ & $8.97 \pm 0.05$ & $0.340 \pm 0.010$ & $0.41 \pm 0.10$ & $46 \pm 5$ & $-3.23 \pm 0.12$ & $-6.28 \pm 0.22$ & $0.51 \pm 0.06$ & -- & Y & 
12, 13, 14, 15
\\
NGC 4472 & $16.72$ & $9.40 \pm 0.10$ & $0.80 \pm 0.09$ & $0.31 \pm 0.13$ & $53 \pm 13$ & $-3.47 \pm 0.21$ & $<-7.62 \pm 0.24$ & $0.66 \pm 0.17$ & --  & Y & 
12, 6, 16, 16
\\
NGC 4486  & $16.68$ & $9.789 \pm 0.027$ & $0.91 \pm 0.11$ & $0.62 \pm 0.05$ & $114 \pm 15$ & $-2.87 \pm 0.08$ & $-5.92 \pm 0.22$ & $1.42 \pm 0.19$ & --  & Y & 
12, 17, 14, 11
\\
NGC 4552 & $15.3$ & $8.92 \pm 0.11$ & $0.350 \pm 0.010$ & $0.6 \pm 0.5$ & $40 \pm 10$ & $-3.1 \pm 0.4$ & $-6.41 \pm 0.24$ & $0.54 \pm 0.14$ & --  & Y & 
5, 18, 14, 15
\\
NGC 4649 & $16.46$ & $9.67 \pm 0.10$ & $1.250 \pm 0.030$ & $0.439 \pm 0.012$ & $63 \pm 14$ & $-3.34 \pm 0.10$ & $-7.93 \pm 0.24$ & $0.80 \pm 0.18$ & --  & Y & 
12, 19, 20, 20
\\
IC 1459 & $28.92$ & $9.39 \pm 0.08$ & $0.56 \pm 0.07$ & $0.31 \pm 0.13$ & $74 \pm 17$ & $-3.25 \pm 0.20$ & $-5.39 \pm 0.23$ & $0.54 \pm 0.12$ & --  & Y & 
5, 21, 22, 22
\\ 
IC 4296 & $46.0$ & $9.43 \pm 0.17$ & $0.560 \pm 0.030$ & $0.97 \pm 0.16$ & $81 \pm 32$ & $-2.72 \pm 0.19$ & $-5.01 \pm 0.28$ & $0.36 \pm 0.14$ & --  & Y & 
24, 18, 23, 25
\\ 
Sgr A$^\ast$ & $0.00828$ & $6.63 \pm 0.04$ & $1.30 \pm 0.20$ & $26 \pm 11$ & $0.056 \pm 0.010$ & $-4.64 \pm 0.22$ & $-8.43 \pm 0.22$ & $1.40 \pm 0.24$ & --  & Y & 
26, 26, 27, 27
\\\hline
NGC 221 & $0.8057$ & $6.39 \pm 0.18$ & $0.37 \pm 0.28$ & $0.069 \pm 0.028$ & $0.11 \pm 0.10$ & $-6.64 \pm 0.31$ & $-7.32 \pm 0.28$ & $0.029 \pm 0.025$ & $17.4$  & N & 
28, 29, 30, 31
\\
NGC 315 & $64.4$ & $9.22 \pm 0.20$ & $0.55 \pm 0.08$ & $0.37 \pm 0.13$ & $51 \pm 25$ & $-3.34 \pm 0.26$ & $-4.44 \pm 0.30$ & $0.16 \pm 0.08$ & $3.04$  & N & 
1, 18, 32, 32
\\
NGC 507 & $70.8$ & $9.21 \pm 0.16$ & $0.490 \pm 0.020$ & $0.8 \pm 0.7$ & $56 \pm 21$ & $-2.9 \pm 0.4$ & $-5.45 \pm 0.27$ & $0.16 \pm 0.06$ & --  & N & 
1, 18, 10, 15
\\
NGC 1291 & $9.08$ & $8.12 \pm 0.16$ & $0.34 \pm 0.05$ & $0.28 \pm 0.11$ & $6.5 \pm 2.6$ & $-4.24 \pm 0.24$ & $-5.54 \pm 0.27$ & $0.15 \pm 0.06$ & $3.35$  & N & 
33, 18, 34, 34
\\
NGC 1316 & $20.95$ & $8.23 \pm 0.08$ & $0.343 \pm 0.007$ & $1.3 \pm 1.2$ & $8.3 \pm 1.5$ & $-3.5 \pm 0.4$ & $-6.07 \pm 0.23$ & $0.082 \pm 0.015$ & --  & N & 
12, 35, 36, 15
\\
NGC 1332 & $22.66$ & $9.17 \pm 0.06$ & $0.60 \pm 0.06$ & $1.0 \pm 0.4$ & $41 \pm 7$ & $-3.03 \pm 0.19$ & $<-7.18 \pm 0.23$ & $0.38 \pm 0.07$ & $1.32$  & N & 
5, 37, 38, 39
\\
NGC 1399 & $20.85$ & $9.10 \pm 0.22$ & $0.80 \pm 0.09$ & $0.44 \pm 0.18$ & $26 \pm 14$ & $-3.62 \pm 0.29$ & $<-6.83 \pm 0.31$ & $0.26 \pm 0.14$ & $1.89$  & N & 
12, 40, 16, 16
\\
NGC 2681 & $17.2$ & $7.53 \pm 0.08$ & $0.67 \pm 0.04$ & $0.43 \pm 0.15$ & $0.86 \pm 0.17$ & $-5.08 \pm 0.18$ & $-5.36 \pm 0.23$ & $0.0104 \pm 0.0020$ & $48.1$  & N &  
5, 18, 32, 32
\\
NGC 4261 & $32.36$ & $8.72 \pm 0.09$ & $0.580 \pm 0.020$ & $1.0 \pm 0.4$ & $15.3 \pm 3.2$ & $-3.45 \pm 0.20$ & $-5.11 \pm 0.23$ & $0.099 \pm 0.020$ & $5.07$  & N & 
5, 41, 14, 39
\\
NGC 4278 & $16.5$ & $8.82 \pm 0.12$ & $0.62 \pm 0.04$ & $1.2 \pm 0.4$ & $18 \pm 5$ & $-3.32 \pm 0.20$ & $-5.25 \pm 0.25$ & $0.22 \pm 0.07$ & $2.23$  & N & 
24, 18, 32, 32
\\
NGC 4438 & $11.61$ & $7.75 \pm 0.06$ & $0.58 \pm 0.10$ & $1.0 \pm 0.4$ & $1.62 \pm 0.35$ & $-4.43 \pm 0.20$ & $-5.76 \pm 0.22$ & $0.029 \pm 0.006$ & $17.2$  & N & 
1, 18, 14, 25
\\
NGC 4594 & $9.87$ & $8.823 \pm 0.027$ & $0.65 \pm 0.34$ & $0.15 \pm 0.06$ & $17 \pm 9$ & $-4.23 \pm 0.28$ & $-5.65 \pm 0.22$ & $0.36 \pm 0.19$ & $1.38$  & N$^{\ast}$ & 
5, 42, 14, 23
\\
NGC 4636 & $14.7$ & $8.49 \pm 0.08$ & $0.60 \pm 0.06$ & $0.11 \pm 0.04$ & $8.6 \pm 1.9$ & $-4.67 \pm 0.20$ & $<-6.13 \pm 0.23$ & $0.122 \pm 0.027$ & $4.11$  & N & 
5, 18, 14, 16
\\
NGC 4696 & $35.5$ & $8.86 \pm 0.15$ & $0.41 \pm 0.04$ & $0.40 \pm 0.25$ & $30 \pm 11$ & $-3.47 \pm 0.31$ & $<-7.58 \pm 0.26$ & $0.17 \pm 0.06$ &  --  &N&
5, 18, 15, 15
\\
NGC 4697 & $12.54$ & $8.31 \pm 0.11$ & $0.330 \pm 0.030$ & $0.019 \pm 0.008$ & $10.3 \pm 2.8$ & $-5.21 \pm 0.21$ & $<-6.55 \pm 0.24$ & $0.17 \pm 0.05$ & $2.93$  & N & 
12, 43, 44, 44
\\
NGC 5005 & $20.0$ & $8.20 \pm 0.09$ & $0.640 \pm 0.030$ & $1.2 \pm 0.4$ & $4.1 \pm 0.9$ & $-3.93 \pm 0.18$ & $-4.90 \pm 0.23$ & $0.043 \pm 0.009$ & $11.6$ & N & 
1, 18, 32, 32
\\
NGC 5044 & $31.2$ & $8.71 \pm 0.17$ & $0.310 \pm 0.010$ & $0.30 \pm 0.10$ & $28 \pm 11$ & $-3.56 \pm 0.22$ & $-6.40 \pm 0.28$ & $0.19 \pm 0.07$ & --  & N & 
5, 18, 15, 15
\\
NGC 5813 & $32.2$ & $8.81 \pm 0.11$ & $0.330 \pm 0.010$ & $0.28 \pm 0.13$ & $33 \pm 8$ & $-3.54 \pm 0.23$ & $-7.09 \pm 0.24$ & $0.21 \pm 0.05$ & --  & N & 
5, 18, 25, 15
\\
NGC 5846 & $24.9$ & $8.82 \pm 0.11$ & $0.378 \pm 0.009$ & $0.40 \pm 0.35$ & $29 \pm 8$ & $-3.5 \pm 0.4$ & $-6.18 \pm 0.24$ & $0.24 \pm 0.06$ & --  & N & 
5, 18, 14, 15
\\
NGC 6166 & $125.3$ & $9.27 \pm 0.12$ & $1.10 \pm 0.07$ & $1.1 \pm 0.7$ & $28 \pm 8$ & $-3.27 \pm 0.30$ & $-5.94 \pm 0.25$ & $0.047 \pm 0.013$ & --  & N &
1, 18, 15, 15
\\
\enddata
\tablecomments{
Column 1: galaxy name; Column 2: distance; Column 3: BH mass; Column 4 and 5: electron temperature and density 
at the Bondi radius; Column 6: Bondi radius; Column 7: Bondi accretion rate normalized by the Eddington rate 
($\dot{m}_{\rm B}\equiv \dot{M}_{\rm B}/\dot{M}_{\rm Edd}$); Column 8: Eddington ratio ($L_{\rm bol}/L_{\rm Edd}$); 
Column 9: angular size of the Bondi radius; Column 10: correction factor of density for unresolved sources;
Column 11: Bondi resolved source = Y (the first 11 sources) and Bondi unresolved source = N (others); 
Column 12: references for the data of ($D$, $M_\bullet$, $L_\mathrm{2-10}$, $T$ \& $n_\mathrm{e}$):
(1) NED, (2) \citealt{Bender_2005}, (3) \citealt{Garcia_2005}, (4) \citealt{Dosaj_2002}, (5) \citealt{Tonry_2001}, (6) \citealt{Rusli_2013}, 
(7) \citealt{Zhang_2004}, (8) \citealt{Humphrey_2006}, (9) \citealt{Emsellem_1999}, (10) \citealt{Ho_2009}, (11) \citealt{Russell_2015},
(12) \citealt{Blakeslee_2009}, (13) \citealt{Walsh_2010}, (14) \citealt{Gonzalez-Martin_2006}, (15) \citealt{Russell_2013}, 
(16) \citealt{Loewenstein_2001}, (17) \citealt{Gebhardt_2011}, (18) \citealt{Kormendy_Ho_2013}, (19) \citealt{Shen_2010},
(20) \citealt{Paggi_2014}, (21) \citealt{Cappellari_2002}, (22) \citealt{Fabbiano_2003}, (23) \citealt{Pellegrini_2003}, 
(24) \citealt{Blakeslee_2001}, (25) \citealt{Pellegrini_2005}, (26) \citealt{Genzel_2010}, (27) \citealt{Baganoff_2003}, 
(28) \citealt{Blakeslee_2010}, (29) \citealt{vandenBosch_2010}, (30) \citealt{Ho_2003}, (31) \citealt{Mathews_2003},
(32) \citealt{Younes_2011}, (33) \citealt{McCourt_2011}, (34) \citealt{Irwin_2002}, (35) \citealt{Nowak_2008}, 
(36) \citealt{Lanz_2010}, (37) \citealt{Rusli_2011}, (38) \citealt{Humphrey_2004}, (39) \citealt{Humphrey_2009},
(40) \citealt{Houghton_2006}, (41) \citealt{Ferrarese_1996}, (42) \citealt{Jardel_2011}, (43) \citealt{Schulze_2011}, and (44) \citep{Soria_2006}.
Note that the angular size of the Bondi radius of the central SMBH in NGC 4594 might be marginally resolved within 
the error, but the spatial resolution of its {\it Chandra} observation reaches only $\simeq 2.5^{\prime\prime}$.
}
\label{tab:1}
\end{deluxetable*}
\end{longrotatetable}

\end{document}